\documentclass{aa}
\usepackage{epsf}
\usepackage{times}
\newcommand{\etal}{{et al}\/.}
\begin{document}
\thesaurus{03{03.13.6; 13.25.2; 11.19.1; 11.17.3; 11.19.3}}
\title{Determining the reality of X-ray filaments}
\author{M.J. Hardcastle}
\institute{Department of Physics, University of Bristol, Tyndall
Avenue, Bristol BS8 1TL, UK (m.hardcastle@bristol.ac.uk)}
\date{Received \today / Accepted \today}
\maketitle
\begin{abstract}
A number of authors have reported filaments connecting bright
structures in high-resolution X-ray images, and in some cases these
have been taken as evidence for a physical connection between the
structures, which might be thought to provide support for a model with
non-cosmological redshifts. In this paper I point out two problems
which are inherent in the interpretation of smoothed photon-limited
data of this kind, and develop some simple techniques for the
assessment of the reality of X-ray filaments, which can be applied to
either simply smoothed or adaptively smoothed data. To illustrate the
usefulness of these techniques, I apply them to archival {\it ROSAT}
observations of galaxies and quasars previously analysed by others. I
show that several reported filamentary structures connecting X-ray
sources are not in fact significantly detected.
\end{abstract}

\section{Introduction}

With the advent of high-resolution X-ray telescopes it is now routine
to see structure in X-ray images. Assessing the level at which one
should believe this structure presents more of a challenge in X-ray
observations than in optical or radio images of comparable resolution,
because X-ray data are very often photon-noise limited. The problem is
made worse by the common (and necessary) practice of smoothing the
data with a Gaussian. This is carried out in order to make the images
conform to our expectations of what the `true' sky image with infinite
exposure time would look like; but it is important to remember that in
general it does {\it not} in fact take away the photon-limited nature
of the underlying data.

A number of smoothed X-ray images of extragalactic objects appear, on
published contour maps, to show quite clear `filamentary' connections
between X-ray sources in the field. (The term `filament' does not
have a single definition in the literature; I shall use it to mean
any extended, apparently linear connection between two sources.) In
this paper I discuss the techniques necessary to determine whether
such filamentary connections are real, and apply them to some sample
observations of galaxies and quasars, previously analysed by other
authors (Arp 1996, Dahlem \etal\ 1996, Arp 1997), which I have taken
from the {\it ROSAT} archives.

\section{Significance and contouring}
\label{significance}

\begin{figure*}
\leavevmode
\begin{center}
\epsfxsize 15cm
\epsfbox{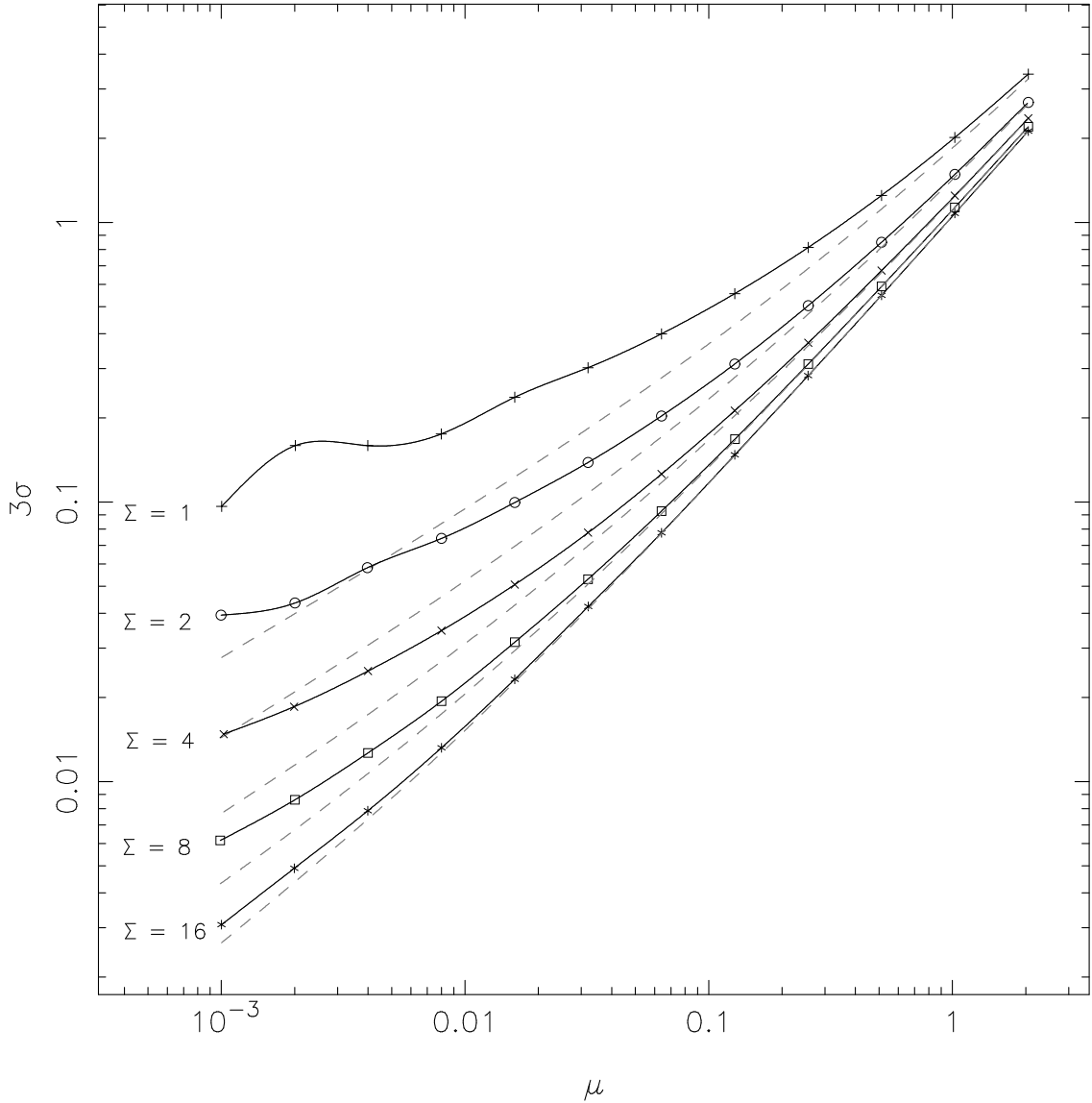}
\end{center}
\caption{The true $3\sigma$ level derived from simulation as a
function of the mean number of counts per pixel in the background
($\mu$) and the width $\Sigma$ of the convolving Gaussian (defined in
the text; FWHM = $2.345\Sigma$). The solid lines are cubic splines fit
through the data points derived from simulation. Deviations from a
smooth curve seen for low $\mu$ and $\Sigma$ are a result of the
Poisson-limited nature of the data. At high $\mu$, it will be seen
that the $3\sigma$ level converges on a value close to $\mu$, as
expected. At low $\mu$ and $\Sigma$, the difference between the
$3\sigma$ level and $\mu$ is highly significant. The dashed lines
below each solid line show the `mean plus three times r.m.s.' contour
levels derived from the same simulations. While they join the true
$3\sigma$ lines at high $\mu$ and $\Sigma$, this quantity is clearly a
very poor, systematically low estimator of the true $3\sigma$ value at
low $\mu$ and $\Sigma$.}
\label{pp}
\end{figure*}

The first problem in assessing the reality of features in smoothed
X-ray maps is that of the significance levels to be chosen when
contouring. If the mean background count density is $\mu$ counts per
unit area and the convolving Gaussian has the form\footnote{To
avoid confusion, I use $\Sigma$ throughout the paper to denote the
width of a convolving Gaussian, and $\sigma$ to refer to the
conventional significance levels.} $f(r) \propto
\exp(-r^2/2\Sigma^2)$, then, as pointed out in Appendix A of Hardcastle
\etal\ (1998), the background noise is {\it not} normally distributed
unless $\mu\Sigma^2 \gg 1$, a condition which does not typically apply
in the sort of datasets being considered here. It is therefore
incorrect, for example, to take the r.m.s.\ dispersion of the
background noise after smoothing, multiply by three, add the mean and
call the resulting contour level `$3\sigma$' as though it were the
corresponding confidence level for normally-distributed noise. In
general it is not. Contours for noise which is not normally
distributed should ideally have the same properties as those for
normally distributed noise, so that the contour referred to as the
`$3\sigma$' contour, for example, should exclude all but 0.135\% of
the background. But in the case where $\mu\Sigma^2 \la 1$, there is
no analytic solution to the question `what is the contour level
corresponding to the $n\sigma$ level for normally distributed noise,
for some $n$?'. Instead, Hardcastle \etal\ (1998) outline a Monte
Carlo procedure for determining the correct significance level. In
this procedure, fields of simulated Poisson noise are convolved with a
Gaussian and the `$n\sigma$' level is derived directly from the
distribution of the resulting noise. (Results of this process for some
sample values of $\mu$ and $\Sigma$ are plotted in Fig.\ \ref{pp}.)
If, on the other hand, $\mu\Sigma^2 \gg 1$, then the noise will
be approximately normally distributed and a simple analysis,
described in Hardcastle \etal\ (1998), shows that the $3\sigma$
contour should be at $\mu + 3\sqrt{\mu/4\pi\Sigma^2}$ counts per
unit area.

It is common for authors not to state the method by which the
`$3\sigma$' (or other) levels in published contour plots of smoothed
X-ray images were derived; in some cases, even the size of the
convolving Gaussian and/or the binning scheme is not given. This
practice makes it very difficult to assess the reality of features in
such plots, and it is recommended that in all cases where the noise is
not, or might not be, normally distributed authors should explain what
they mean by `$3\sigma$'. In what follows, wherever I refer to a
$2\sigma$ or $3\sigma$ level calculated by me, it will imply that I
have derived it from simulated Poisson noise as described above.

\section{The effect of the PSF}
\label{psf}

However, even when the $3\sigma$ level is correctly defined, such that
a contour at this level would exclude all but 0.135\% of the
Gaussian-smoothed background noise, it is unfortunately not
necessarily the case that we can say that all `structure' around a
bright source enclosed by such a contour is detected at a
corresponding significance level. The radially averaged point-spread
function (PSF) of X-ray instruments often has a very broad tail, so
that there are expected to be significant count densities at large
distances from the centroid; the shape of the PSF can be
energy-dependent, so that these off-centre counts are likely to have a
peculiar spectrum compared to either the point source itself or the
background. When two point sources are placed close to each other, the
natural consequence of this feature of the PSF is that we get an
enhancement of the count rate which is at its greatest directly
between them. Smoothing the data with a Gaussian may well bring this
feature up above the noise, giving the appearance of a filament
connecting the sources.
\begin{figure*}
\leavevmode
\begin{center}
\epsfxsize 15cm
\epsfbox{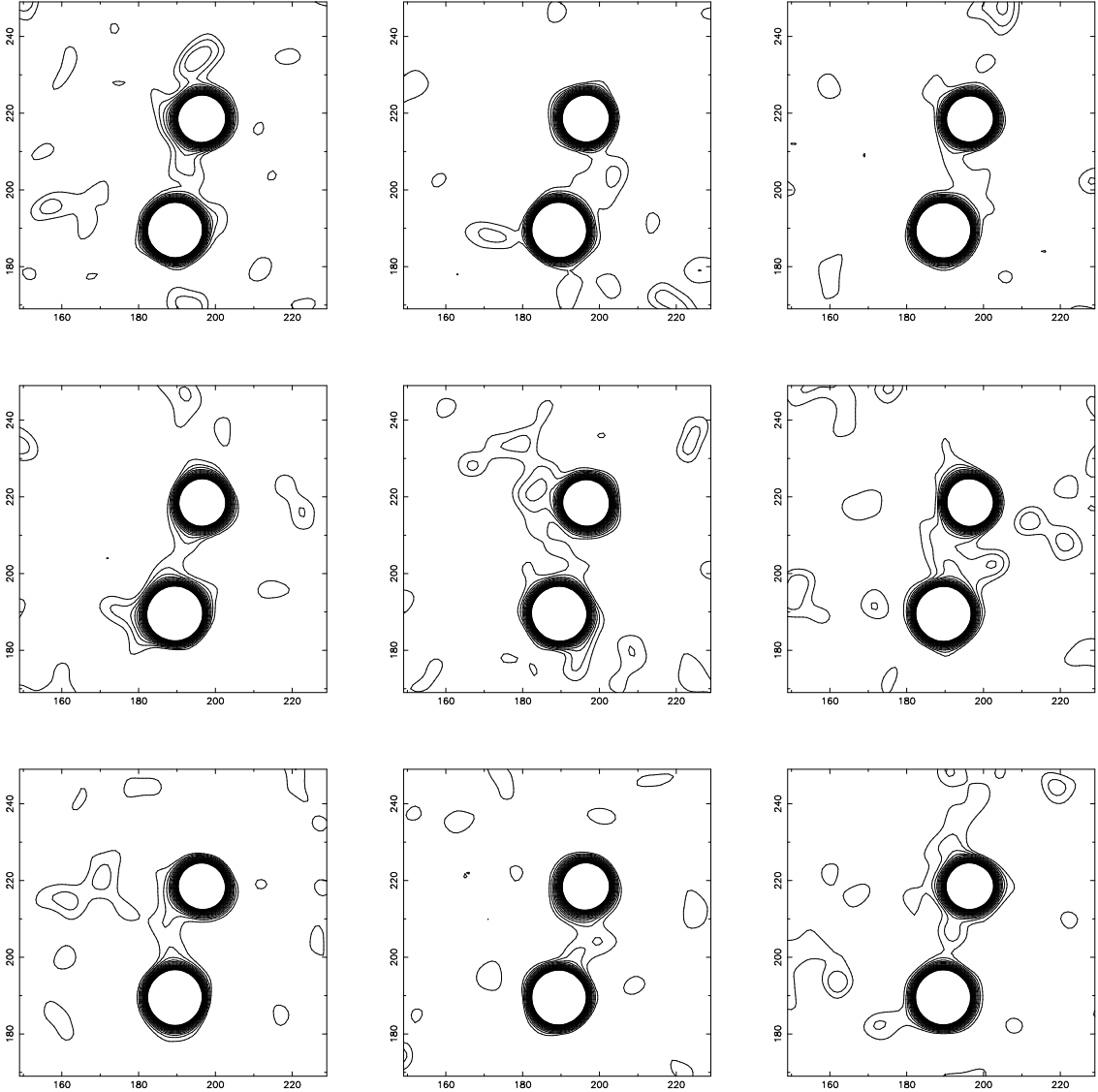}
\end{center}
\caption{Simulated {\it ROSAT} HRI observations of the Seyfert
NGC 4151 and a neighbouring BL Lac object, binned in 10-arcsec pixels
and smoothed with a $\Sigma = 3$ pixel Gaussian. At low contour
levels, a variety of apparent filamentary structures of the type shown
can be seen connecting the sources in about one in five simulated
observations. The actual observations of this object are discussed in
section \ref{ngc4151dis}.}
\label{simfilaments}
\end{figure*}

So to assess the reality of filaments observed at a given contour
level in smoothed images we must ask what is the probability of seeing
such filaments under the null hypothesis, which is that the two
sources are in fact not connected by any real feature on the X-ray
sky. In general this too must be done by Monte Carlo simulation. I
have written code which generates a Poisson-distributed noise
background and two or more simulated sources, convolved with a
suitable PSF, which match those seen on the sky. The simulated data
are then smoothed with a Gaussian, and then either automatically or by
eye it is possible to see whether a `filament' at a given contour
level joins them (in practice, the code does this using a simple
recursive flood-fill algorithm). By repeating this procedure many
times, we can directly estimate the probability that an observed
filament would be seen under the null hypothesis, and so
characterise its statistical significance. Only connections which have
a very small chance of being seen under the null hypothesis can be
described as statistically significant. An example of the kinds of
apparent filaments that can result from low choices of contour levels
even under the null hypothesis is given in Fig.\ \ref{simfilaments}.

\section{Data and simulations}

\begin{table}
\caption{{\it ROSAT} observations obtained from the public
archive. The ROR number is the {\it ROSAT} Observation Request; an `h'
as the second letter denotes HRI data, a `p' indicates PSPC.}
\label{data}
\begin{tabular}{llr}
Object&ROR number&Livetime (s)\\
\hline
Mrk 474&rp600448&12381\\
NGC 4651/3C\,275.1&rp600450&10147\\
&rh800719&25157\\
NGC 3067/3C\,232&rp700468&5376\\
NGC 4319/Mrk 205&rh600441&12272\\
&rh600834&34991\\
NGC 3628&rp700010&13387$^*$\\
&rh700009&13484\\
NGC 4151&rh700023&9162\\
\end{tabular}
\vskip 8pt
\begin{minipage}\linewidth
$^*$ After master veto rate filtering; see the text.
\end{minipage}
\end{table}

Using the techniques discussed above, it is possible to re-examine
objects discussed in the literature. The {\it ROSAT} data
listed in Table \ref{data} were obtained from the public archive.
Data were analysed using the IRAF Post-Reduction Offline Software (PROS).

The radially averaged {\it ROSAT} PSPC PSF is well known (e.g.\
Hasinger \etal\ 1995) and so, using a suitably integrated version of
an analytical approximation to the PSF, it is simple to simulate PSPC
observations. One complication is that the PSF is both
energy-dependent and dependent on the off-axis angle of the
source. Dealing with the energy dependence of the PSF is not hard, but
would involve generating an energy-weighted PSF for each source. For
simplicity, I choose to use the PSF at a single energy, in most cases
1 keV (the energy at which {\it ROSAT} is most sensitive); this is a
conservative choice, in the sense that the PSF is narrowest at
approximately this energy. The off-axis angle dependence of the width
of the PSF is incorporated, though not the distortions which appear at
far-off-axis angles; no source considered here is significantly
affected by them. The actual PSF observed in a given observation is
also affected by the {\it ROSAT} aspect reconstruction problems, which
can introduce various kinds of elongation, but since it is very hard
to determine the extent to which this affects a given observation, I
have not included it in the simulations.

The {\it ROSAT} HRI PRF is more strongly affected by the aspect
reconstruction problems, but, for simplicity, in the two cases where it
is used in this paper I have just used the standard parametrization
of David \etal\ (1997).

\subsection{Mrk 474}

Mrk 474 is a $z=0.041$ Seyfert galaxy adjacent to a $z=1.94$ quasar
(Arp 1996). The Seyfert is a very strong X-ray source, while the point
X-ray source identified by Arp with the quasar, located 170 arcsec to
the NW, is only weakly detected in the PSPC observations.

\begin{figure}
\epsfxsize\linewidth
\epsfbox{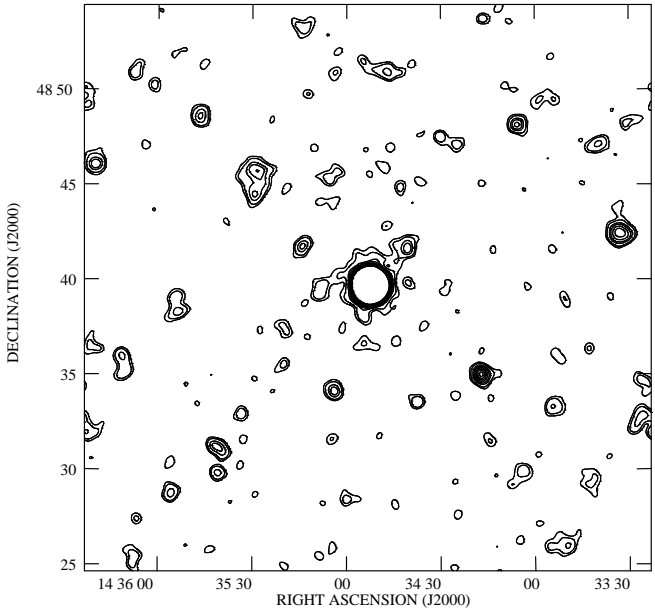}
\caption{The {\it ROSAT} PSPC image of Markarian 474 smoothed and
binned as described in the text, with contours at $0.0319, 0.0435,
0.07, 0.09, 0.12 \dots 0.30$ counts pixel$^{-1}$. The lowest contours
correspond to the $2$ and $3\sigma$ levels derived from simulation.}
\label{mrk474}
\end{figure}

Arp (1996) analysed the 0.5-2.0 keV PSPC data, and saw a filament
connecting the Seyfert and quasar. It is not clear what binning scheme
and smoothing Gaussian was used to produce his figures 1 and 2, but I
obtain very similar results by binning by a factor 7 (so that each
pixel is 3.5 arcsec) and smoothing with a Gaussian of $\Sigma = 4$
pixels (14 arcsec). The results are insensitive to the exact values
of binning and smoothing used. The mean number of background counts in
the image per 3.5-arcsec pixel, close to the central source and after
excluding all visible background sources is 0.0135, which, using the
techniques described in section \ref{significance}, corresponds to
$2$- and $3\sigma$ levels in the smoothed dataset of 0.032 and 0.044
counts pixel$^{-1}$ respectively. The image is contoured with these
levels in Fig.\ \ref{mrk474}. Comparing this figure with Arp's figures
1 and 2, it can be seen that my $2\sigma$ level is more or less
equivalent to his $4\sigma$ level, while my $3\sigma$ level is close
to his $5\sigma$ level. As shown in Fig.\ \ref{mrk474}, the connecting
`filament' is not represented except at the $2\sigma$ level when the
confidence limits are calculated correctly. Moreover, when I simulated
this situation (using 30 net counts for the quasar, as measured by Arp
(1996), and 6700 net counts for the Seyfert galaxy), I found that we
expect to see a connection between the two sources at the $2\sigma$
level 65\% of the time, and even at the $3\sigma$ level 30\%
of the time, if the two sources are in fact just adjacent point
sources, because of the increased count density close to Mrk 474. The
detection of the filament in this source is thus certainly not
statistically significant.

\subsection{NGC 4651/3C\,275.1}

The nearby (Virgo cluster) galaxy NGC 4651 is adjacent to the
radio-loud quasar 3C\,275.1 ($z=0.557$). In this case the quasar is a
reasonably strong X-ray source, while the galaxy shows some weaker
extended emission. Hardcastle \& Worrall (1999) found the PSPC
emission from the quasar to be consistent with that of a point source,
although the HRI observations of the source suggest some extension on
small scales, consistent with emission from a cluster (Hardcastle \&
Worrall 1999, Crawford \etal\ 1999). NGC 4651 has an optical jet or
tail; Arp (1996) views this as evidence for ejection, while Wehrle,
Keel \& Jones (1997) suggest that the optical colours are evidence for
a tidal origin.

Binning and smoothing the 0.1-2.0 keV energy band with the same
parameters as I used for Mrk 474, and with a measured mean background
count level of 0.0534 counts pixel$^{-1}$, the elongation of NGC 4651
in the direction of 3C\,275.1 is clearly visible. The $3\sigma$
contour after smoothing is at about 0.110 counts pixel$^{-1}$, and the
elongation is seen at this level [corresponding to a level somewhere
between the $4$ and $8\sigma$ contours in figure 7 of Arp
(1996)]. Taking the quasar as containing about 200 counts in the
0.1-2.0 keV energy band, and the galaxy as containing 120, I used
simulations to determine whether, if the galaxy were really a point
source, the effect of the quasar's proximity could cause the observed
extension; as expected from the images and from the comparative
weakness of the quasar, the probability of obtaining such a strong
result by chance is negligible. We can safely conclude that the
extension of NGC 4651 is likely to be real. However, there is no
evidence for a compact jet-like structure in NGC 4651 in the HRI
image. 

\subsection{NGC 3067/3C\,232}

The radio-loud quasar 3C\,232 ($z=0.534$) is adjacent to the starburst
galaxy NGC 3067. In this case, too, the effect of contamination by the
quasar's PSF is small and the elongation seen by Arp (1996) in the
0.7--2.0 keV X-ray emission of NGC 3067 appears real; simulation shows
that this degree of extension in the smoothed image is too great to be
produced by chance, particularly as it is not directed towards the
quasar (which is what we would expect for an artefact of the sort
discussed in section \ref{psf}).

Arp (1996) also reports filaments in smoothed 0.1-2.0 keV images
connecting 3C\,232 to stellar objects to the north and northeast. Arp
binned the data here in 5-arcsec pixels and smoothed with a small
Gaussian of $\Sigma=3$ pixels. The background noise level per 5-arcsec
pixel in this energy band is 0.068 counts; the corresponding $3\sigma$
level is 0.15 counts pixel$^{-1}$, whereas Arp's lowest contour
appears to have been at 0.11 counts pixel$^{-1}$, somewhat below the
$2\sigma$ level.  Simulation shows that we would not expect a filament
even at this level connecting 3C\,232 with the northern stellar object
by chance under the null hypothesis, which suggests that there is some
real emission in between 3C\,232 and the northern stellar object, but
the data provide no significant evidence for such a filament rather
than, for example, an intervening point source.

\subsection{NGC 4319/Mrk 205}

Arp (1996) observed the X-ray bright $z=0.07$ Seyfert galaxy Mrk 205
with the {\it ROSAT} HRI. The Seyfert lies 44 arcsec south of the
galaxy NGC 4319, which was not detected in the HRI observations; but a
number of other sources are detected in the field, including the
early-type galaxy NGC 4291, 6.4 arcmin away to the NE, which is
extended in the X-ray. Three of the brightest of these sources, at
off-axis distances from 8 to 15 arcmin, are identified with
intermediate-to-high redshift quasars. In addition to Arp's data, a
deeper exposure centred on NGC 4291 exists in the archive; this shows
similar detections. Having adaptively smoothed his dataset (this
process involves smoothing the pixels containing small numbers of
photons with larger Gaussians than those containing larger numbers,
and then summing the results) Arp obtained the `startling' result
that many --- indeed, almost all --- of the point sources were joined
to the central source, Mrk 205, by `filaments' of X-ray emission.

I began to investigate this field by analysing the data in a similar
way to the other sources. Adopting Arp's binning scheme of 8-arcsec
pixels, the mean number of background counts per pixel $\mu$ is then
about 1.08. I smoothed the whole image with a Gaussian of 16 pixel
FWHM, or $\Sigma = 6.80$ pixels; the corresponding $3\sigma$ level is
1.21 counts pixel$^{-1}$. (This result was derived from numerical
simulation, but since $\Sigma^2\mu \gg 1$ in this case, the background
noise is close to being normally distributed, and the analytic
approximation described in section \ref{significance} in fact applies and gives
the same answer.) At this level, or even at the corresponding
$2\sigma$ level, no filaments connected any of the sources. Only at
the $1\sigma$ contour level did the filamentary structure seen in
Arp's figure 10 begin to become apparent. 

However, Arp's use of adaptive smoothing is intended to improve the
visibility of low-surface-brightness features. The characteristics of
Poisson noise smoothed in this way are even less clear than for the
single-Gaussian case, but the question may still be answered by Monte
Carlo simulation as discussed in section \ref{significance}. Arp's
description of his binning scheme is somewhat ambiguous, but I
obtained a reasonable match to the image in his figure 10 by smoothing
the 8-arcsec pixels containing one count with a Gaussian of $\Sigma$
(not FWHM) of 16 pixels, pixels with two counts with a Gaussian of
$\Sigma = 16/\sqrt{2}$ pixels, and so on ($\Sigma = 16 \times
2^{-(n-1)/2}$ pixels, where $n$ is the number of counts) up to pixels
with more than 12 counts, which were not smoothed.

\begin{figure}
\epsfxsize\linewidth
\epsfbox{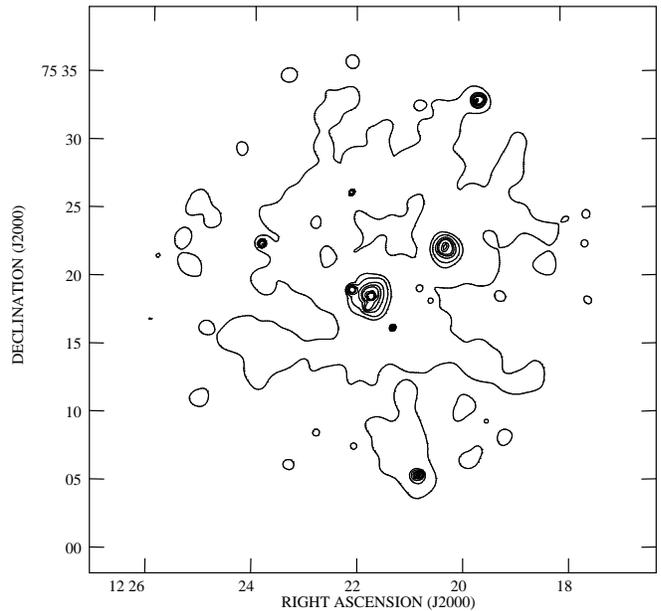}
\caption{The {\it ROSAT} HRI Markarian 205 field adaptively smoothed
as described in the text. The lowest contour, 1.05 counts
pixel$^{-1}$, shows up non-significant structure comparable to that in
the image of Arp (1996). The next contour, 1.29 counts pixel$^{-1}$,
is the $3\sigma$ level derived from simulation. This image shows the
whole HRI field of view, which is approximately square with the
vertices lying on the axes of the plot to the N, S, E and W of Mrk
205. (Note that some detected but faint point sources are not enclosed by
the $3\sigma$ contour; this is because adaptive smoothing is
optimized to emphasize low-surface-brightness extended emission,
and tends to smear out the counts from weak point sources.)}
\label{mrk205}
\end{figure}

I then simulated the effects of smoothing Poisson-distributed noise
with $\mu = 1.08$ counts pixel$^{-1}$ with this adaptive filter. The
resulting $3\sigma$-equivalent level (1.29 counts pixel$^{-1}$) is, as
expected, higher for this smoothing method than for the case where the
whole field is smoothed with the largest Gaussian (on the analytic
approximation, this would be 1.13 counts pixel$^{-1}$), both because
the `average' smoothing Gaussian is smaller and because the method
accentuates the difference between bright and faint noise pixels. As
shown in Fig.\ \ref{mrk205}, when contoured at this level my
adaptively smoothed image shows no significant filaments; a much lower
level of contouring is required to show up the structures in Arp's
figure 10. In fact, it appears from the caption from Arp's figure 11a
that his `$3\sigma$' level is 1.08 counts pixel$^{-1}$, which is
similar to the {\it expected number} of counts per pixel (unchanged,
of course, as a result of smoothing). It is therefore not at all
surprising that contouring at this level shows connections between
adjacent sources; since the median for these data is close to the
mean, this contour level corresponds to contouring about half the area
of the map (preferentially the central regions, because of the effects
of vignetting on the external component of the X-ray
background). Simulating adaptive smoothing of the HRI field of view,
neglecting vignetting, I find that we expect to see a connection at
this level between the central source and the northernmost point
source (identified with a $z=0.464$ quasar) approximately 25\%
of the time under the null hypothesis; the filaments in Arp (1996) are
therefore not statistically significant. They are similarly undetected
on the deeper HRI image centred on NGC 4291 (where the $3\sigma$ level
of the adaptively smoothed image is lower as a fraction of $\mu$
because of the higher number of background counts per pixel).

\subsection{NGC 3628}

The starburst galaxy NGC 3628 was observed by Dahlem \etal\ (1996) and
shows a striking filamentary structure, extending south from the
galactic nucleus to a point source 2.7 arcmin away, in images in the
0.75 keV band (0.44--1.21 keV) and 1.5 keV band (0.73--2.04 keV); this
point source itself appears connected to another point source a short
distance to the SE, which Flesch \& Arp (1999) identify with a
$z=2.15$ quasar.

The contour levels selected by Dahlem \etal\ correspond quite well to
the $3\sigma$ levels derived from simulation; after master veto rate
filtering to match theirs and binning in 15-arcsec pixels I find a
background count rate away from NGC 3628 in the 0.75-keV band of 0.23
counts pixel$^{-1}$, corresponding to a $3\sigma$ level of 0.73 counts
pixel$^{-1}$, which is actually below their adopted $2.5\sigma$
level. The filament is thus apparent in contours at better than the
$3\sigma$ level. Simulation (using a PSF appropriate for an energy of
0.75 keV) shows that the connection between the two adjacent point
sources is not significant, but that a connection at the observed
level between NGC 3628 and the source to the S is not expected to
occur by chance. However, this does not take account of the extended
X-ray halo found by Dahlem \etal\ (1996) around NGC 3628 (and clearly
visible as excess emission in their figure 2). If we model this
crudely as an extra uniform contribution to the background --- taking
the average in a region between NGC 3628 and the southern source, the
background is pushed up to $0.59 \pm 0.04$ counts pixel$^{-1}$ ---
then the connection between NGC 3628 and the point source can occur by
chance in a few per cent of cases, so the feature is only marginally
significant. Again, the filamentary emission is not seen in the HRI
image, which supports the notion that we are seeing a region of
extended emission rather than a true X-ray jet; but it will clearly be
of great interest to obtain more sensitive images of this field. At
the time of writing, a scheduled short {\it Chandra} observation has
not yet been processed.

\subsection{NGC 4151}

\begin{figure}
\epsfxsize\linewidth
\epsfbox{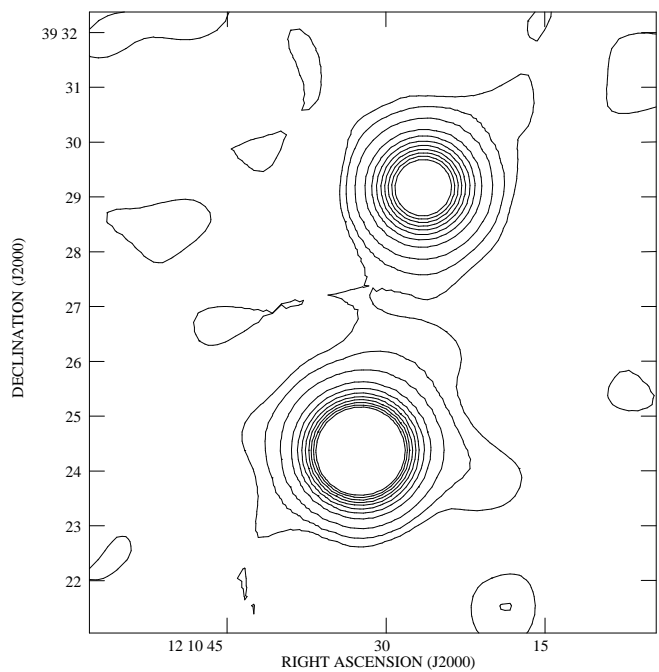}
\caption{The {\it ROSAT} HRI NGC 4151 field smoothed and
binned as described in the text, with contours at $1.35, 1.53, 2, 3,
4\dots 10$ counts pixel$^{-1}$. The lowest contour shows up
non-significant structure at the level of the features seen in figure
17 of Arp (1997). The next contour is the $3\sigma$ level derived from
simulation. This image may be compared to the simulated images of
Fig.\ \ref{simfilaments}.}
\label{ngc4151}
\end{figure}

Arp (1997) suggests that a filament connects this well-studied
Seyfert galaxy to a $z=0.615$ BL Lac object 5 arcmin to the N in a
{\it ROSAT} HRI image. Arp binned the HRI data into 10-arcsec
(20-pixel) square bins; the off-source background is then 1.21 counts
pixel$^{-1}$. It is not clear what Arp means when he says that he
smoothed the data with a `$5\sigma$ Gaussian'; a $\Sigma = 5$ arcsec
Gaussian is very small given this binning scheme while a $\Sigma = 5$
pixel Gaussian is very large, causing the two sources to overlap. I
obtain images quite similar to those of Arp (1997) if I smooth with a
$\Sigma = 3$ pixel Gaussian (Fig.\ \ref{ngc4151}). The $3\sigma$ level
derived from simulation is then 1.53 counts pixel$^{-1}$ (since
$\mu\Sigma^2 \approx 10$, the analytic approximation of section
\ref{significance} gives a very similar answer). As Fig.\
\ref{ngc4151} shows, the sources are not connected at the $3\sigma$
level with this choice of smoothing. A connection between the sources
does appear at a much lower significance level, 1.35 counts
pixel$^{-1}$, which is about $1.5\sigma$. Simulation shows that a
connection at this level can occur by chance 20\% of the time under
the null hypothesis (some examples of the connections seen in
simulations are shown in Fig.\ \ref{simfilaments}, where the lowest
contour is at 1.35 counts pixel$^{-1}$). The connection seen in Fig.\
\ref{ngc4151} is therefore not significant.
\label{ngc4151dis}

\section{Conclusions}

Using Monte Carlo techniques to find the distribution of pixel values
in smoothed noise fields and to assess the reality of connections
between adjacent sources, I have found that the filamentary X-ray
connections between low- and high-redshift Seyfert galaxies and
quasars reported by Arp (1996) are not statistically significant. I
have confirmed that the X-ray extension of two low-redshift galaxies
located near high-redshift radio-loud quasars, reported in the same
paper, is significant. The filament extending south from NGC 3628,
discussed by Dahlem \etal\ (1996), was shown to be marginally
significant. The connection between NGC 4151 and a BL Lac object
reported by Arp (1997) is also apparently not statistically significant.

These results suggest that it is necessary to be cautious in the
interpretation of smoothed X-ray images, and I urge authors to apply
the techniques I have described to any situation where the detection
or non-detection of faint extended X-ray emission has important
scientific implications.

\begin{acknowledgements}
I am grateful to Eric Flesch for kindly arranging for me to be sent
the copy of `Seeing Red' (Arp 1998) which provoked my interest in this
topic, and to two referees for their helpful suggestions.
\end{acknowledgements}

\end{document}